\documentclass[aps,pre,twocolumn,superscriptaddress,showkeys,amsmath,amssymb]{revtex4}
\usepackage[unicode,pdfstartview=FitH,bookmarksnumbered,breaklinks,colorlinks,linktocpage,citecolor=blue,linkcolor=red,urlcolor=blue]{hyperref}

\usepackage{epsfig,amsmath,amssymb,color}
\usepackage[T1]{fontenc}
\usepackage[latin9]{inputenc}

\begin{document}

\title{One-dimensionally confined ammonia molecules: A theoretical study}

\author{Maksym~Druchok}
\email{maksym@icmp.lviv.ua}
\affiliation{Institute for Condensed Matter Physics,
         National Academy of Sciences of Ukraine,
         Svientsitskii Street 1, 79011 L'viv, Ukraine}

\author{Volodymyr~Krasnov}
\email{krasnoff@icmp.lviv.ua}
\affiliation{Institute for Condensed Matter Physics,
          National Academy of Sciences of Ukraine,
          Svientsitskii Street 1, 79011 L'viv, Ukraine}

\author{Taras~Krokhmalskii}
\email{krokhm@icmp.lviv.ua}
\affiliation{Institute for Condensed Matter Physics,
          National Academy of Sciences of Ukraine,
          Svientsitskii Street 1, 79011 L'viv, Ukraine}

\author{Oleg~Derzhko}
\email{derzhko@icmp.lviv.ua}
\affiliation{Institute for Condensed Matter Physics,
          National Academy of Sciences of Ukraine,
          Svientsitskii Street 1, 79011 L'viv, Ukraine}
\affiliation{Professor Ivan Vakarchuk Department for Theoretical Physics,
          Ivan Franko National University of L'viv,
          Drahomanov Street 12, 79005 L'viv, Ukraine}

\date{\today}

\begin{abstract}
We examine a single-file chain of ammonia molecules in a carbon nanotube.
To this end,
we use
i) molecular dynamics simulations (combined with the charges for ammonia nitrogen and hydrogen obtained from quantum chemistry)
and
ii) lattice-model calculations [M.~Druchok {\it et al.}, J. Chem. Phys. {\bf 158}, 104304 (2023)].
Our findings demonstrate the occurrence of the orientational quasiorder of the ammonia dipoles,
which become parallel to the tube axis,
at intermediate temperatures below $100$~K.
\end{abstract}


\maketitle

The authors have a great pleasure to dedicate this paper to the upcoming eighties birthday of Professor M.~F.~Holovko,
whose broad research activity, inspiring talks and lectures, relevant comments at various occasions, and simply good advises
have strong influence on a few generations of researches in statistical physics and soft matter theory in L'viv.

\section{Introductory remarks}
\label{sec1}

Confinement of molecules to pores of nanometer diameters,
when they form a single-file chain,
results in essentially new properties of such a substance.
Single-walled carbon nanotubes~(CNTs) of a corresponding diameter
provide an excellent experimental setup to study the one-dimensionally confined substance~\cite{Cambre2010}.

Recently, X.~Ma {\it et al.}~\cite{Ma2017} have reported temperature-dependent photoluminescence spectroscopy data for single-walled (6,5) CNTs.
While empty CNTs exhibit a linear temperature-dependent photoluminescence spectral shift as expected,
the water-filled CNTs show a stepwise photoluminescence spectral shift centered at about $150$~K,
which is superimposed on the anticipated linear temperature-dependent one.
X.~Ma {\it et al.}~\cite{Ma2017} assumed that the origin of the observed additional spectral shift
is related to a significant change in the orientation of the water dipoles.
They performed molecular dynamics (MD) simulations and indicated three different regimes:
1) traditional hydrogen-bonded chains (below $\sim 40$~K),
2) predominantly bifurcated hydrogen bonds,
where hydrogen bond from a single oxygen atom is distributed over both hydrogen atoms of a single neighboring water molecule (around $\sim 70$~K),
and
3) disordered chains (for $T>200$~K).
The effective total dipole moment of the structures dominated as temperature grows
agrees with the direction of the measured photoluminescence spectral shift.
Several theoretical studies have been inspired by experiments reported in Ref.~\cite{Ma2017}.
Thus, the ground-state and finite-temperature properties of one-dimensionally confined water molecules
were discussed in Refs.~\cite{Sahoo2021,Serwatka2022a,Serwatka2022b,Serwatka2023,Druchok2023}.

In the present study,
we address a question whether quasiphase transitions, observed for the water molecules H$_2$O encapsulated in single-walled (6,5) CNT,
can be expected for other molecules.
To this end, we consider the ammonia molecules NH$_3$
and the same, yet fillable, CNT.
Ammonia molecule has a dipole moment that makes it polar and it has ability to form hydrogen bonds~\cite{Bauer2007}.
Similarly to Refs.~\cite{Ma2017,Druchok2023}, we carry out
i) quantum chemistry calculations to obtain charges of H and N inside the (6,5) CNT,
ii) MD simulations to obtain the dipole moment dependence on temperature for illustration of quasiphases
as well as to obtain reference values for a lattice model
(dipole moment and intermolecular distance),
and
iii) lattice-model calculations.
Our theoretical analysis gives evidence
that the ammonia molecules encapsulated in the (6,5) CNT may show a temperature-driven orientational quasiordering
at intermediate temperatures below 100~K.

The rest of the paper is organized as follows.
In Section~\ref{sec2}, quantum-chemical computations and MD simulations are described.
In Section~\ref{sec3}, we report the lattice-model calculations to further illustrate a temperature-driven dipole quasiordering.
Then, we conclude with a brief discussion and summary in Section~\ref{sec4}.

\section{Molecular simulations}
\label{sec2}

\subsection{Quantum chemistry and ammonia molecule model}
\label{sec21}

To examine the ammonia molecules encapsulated in (6,5) CNT by MD simulations,
the realistic charges for the ammonia nitrogen and hydrogen atoms are primarily required.
According to Ref.~\cite{Eckl2008},
for the bulk ammonia case
$q_{\rm N}=-0.9993e$ and $q_{\rm H}=-q_{\rm N}/3$, where $e$ is the elementary electric charge.
The OPLS~(Optimized Potentials for Liquid Simulations) force field~\cite{Jorgensen1996}
gives $q_{\rm N}=-1.026e$ and $q_{\rm H}=-q_{\rm N}/3$.
However, these charge values do not account the presence of CNT
and therefore are of limited applicability for MD simulations of ammonia molecules inside CNT.

\begin{figure}[htb!]
\begin{center}
\includegraphics[clip=true,angle=0,width=0.7\columnwidth]{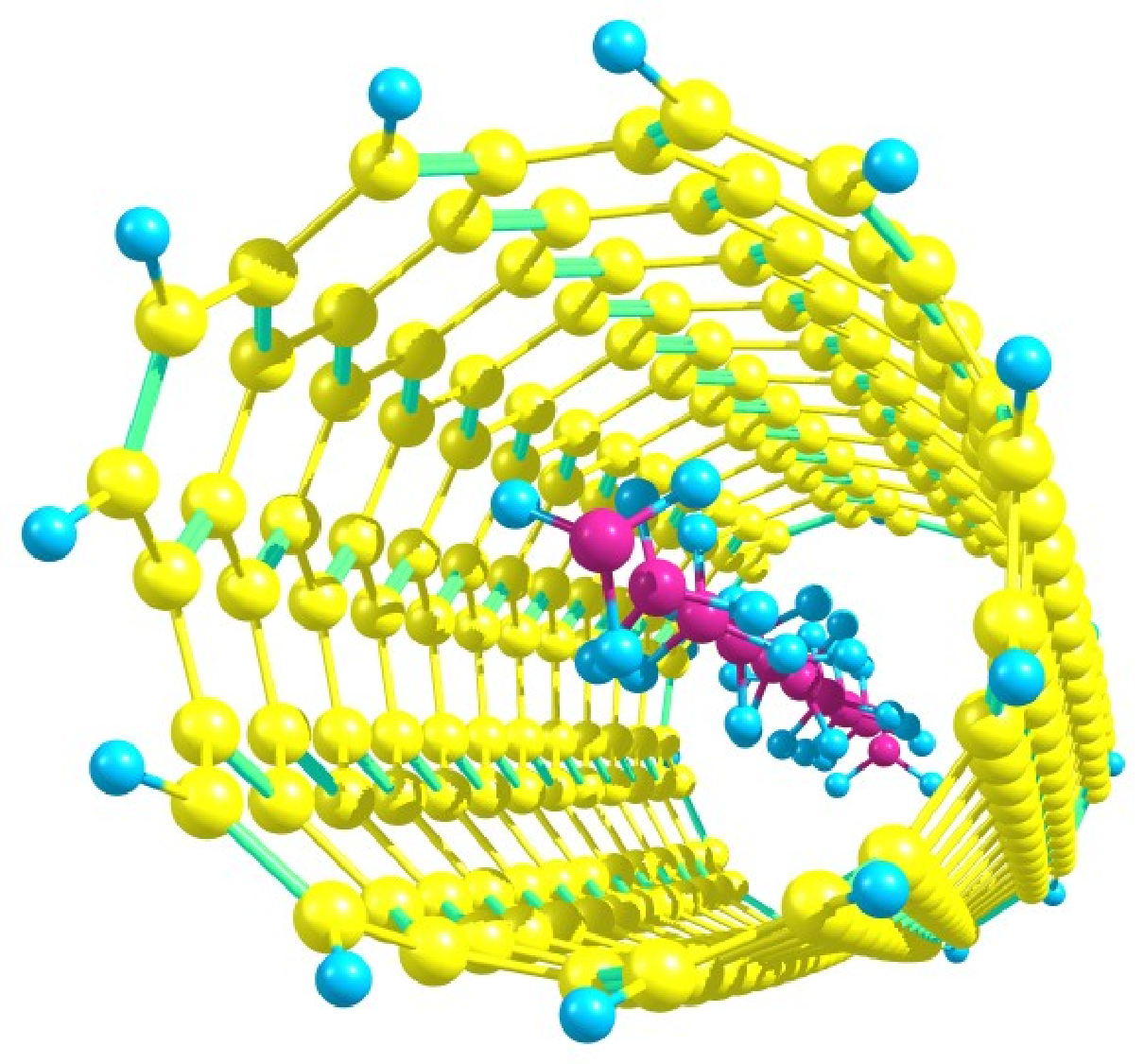}\\
\vspace{3mm}
\includegraphics[clip=true,angle=0,width=1.0\columnwidth]{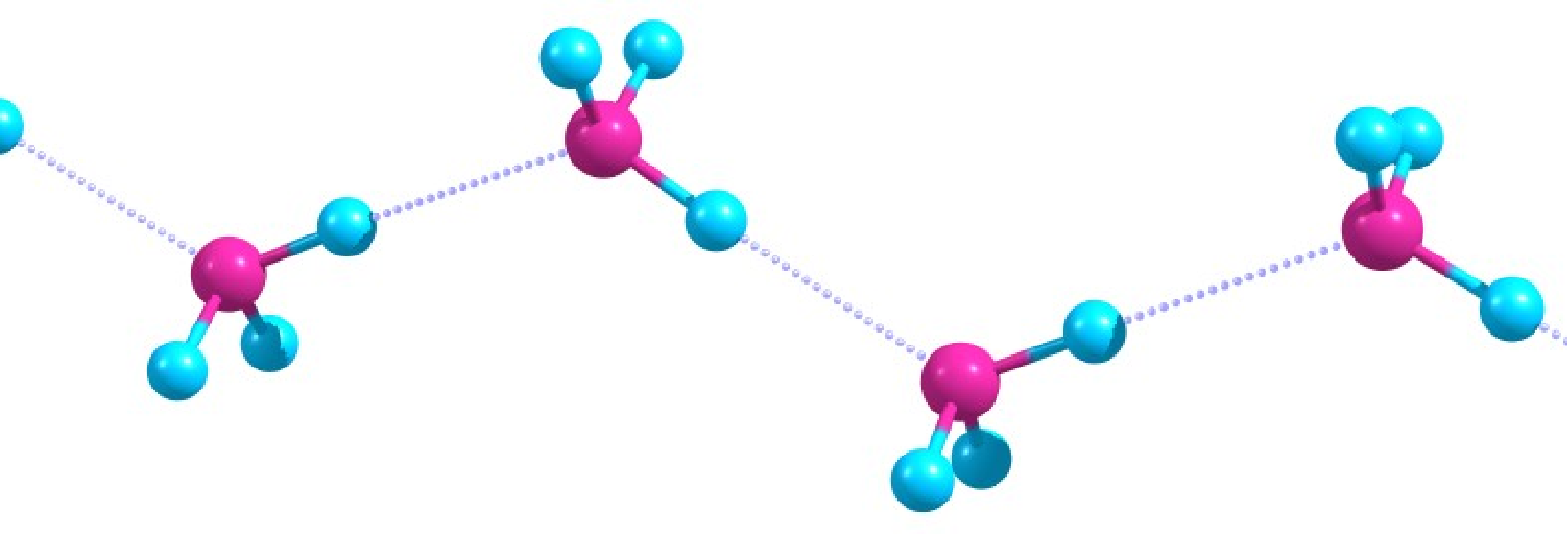}\\
\caption{Quantum chemistry predictions for one-dimensionally confined $N=10$ ammonia molecules.
Top panel: Visualization of AM1 data.
Bottom panel: Visualization of B3LYP data (only the 4th, 5th, 6th, and 7th ammonia molecules are shown).}
\label{fig01}
\end{center}
\end{figure}

To obtain the charges for the nitrogen and hydrogen atoms in ammonia in the (6,5) CNT, we used the GAMESS package~\cite{Schmidt1993}
and performed semi-empirical (AM1, PM6), first-principle (STO-2G, MINI), and density-functional-theory (B3LYP) calculations.
Within these calculations,
we examined
ammonia molecules inside a (6,5) CNT (AM1, PM6 -- see the top panel of Fig.~\ref{fig01})
and
ammonia molecules restricted to one dimension but without CNT (AM1, PM6, STO-2G, MINI, B3LYP -- see the bottom panel of Fig.~\ref{fig01}).
To model the (6,5) CNT, we used the structure with 272 carbon atoms and 22 hydrogen atoms added to saturate free carbon bonds on the edges of CNT
(see the top panel of Fig.~\ref{fig01}).
The coordinates of carbon atoms were frozen while for the coordinates of hydrogen atoms the optimal positions were found.
In our study, we consider 10 ammonia molecules and search for optimal positions of the nitrogen and hydrogen atoms.
To avoid surface effects, i.e., to minimize the effects of molecules at terminal positions,
only charges of the four inner ammonia molecules were averaged for further molecular dynamics simulations.

Mean charges obtained at the AM1/PM6 level are:
$q_{\rm N}=-0.3965/-0.6127$ (in units of $e$) for nitrogen atoms and $q_{\rm H}=0.1322/0.2042$ for hydrogen atoms.
In the presence of CNT we obtained $q_{\rm N}=-0.3913/-0.6281$ and $q_{\rm H}=0.1289/0.2100$ at the AM1/PM6 level, respectively.
Only small changes in atomic charges allow us to examine the former case alone while performing more demanding density-functional-theory calculations.
Furthermore,
the B3LYP method yields
$q_{\rm N}=-0.7814$ (L\"{o}wdin population analysis)
or
$q_{\rm N}=-0.9582$ (Mulliken population analysis)
for nitrogen atom
and
$q_{\rm H}=0.2704/0.2540$ (L\"{o}wdin)
or
$q_{\rm H}=0.3761/0.2918$ (Mulliken)
for hydrogen atoms.
Here, doubled values of $q_{\rm H}$ correspond to
the charge of hydrogens forming hydrogen bonds (higher value before the slash)
and
the charge of two dangling hydrogens (lower value after the slash),
see the bottom panel in Fig.~\ref{fig01}.
Semi-empirical AM1/PM6 calculations also provide some hints for forming the hydrogen bonds,
however, the difference in two values of charges $q_{\rm H}$ is smaller.
It should be stressed that the values of $q_{\rm N}$ (and $q_{\rm H}$) are lower than the bulk ones,
found in Ref.~\cite{Eckl2008} to fit to experimental vapor-liquid equilibrium data ($q_{\rm N}=-0.9993$) or from OPLS ($q_{\rm N}=-1.026$).

Lower charges result in a reduced dipole moment of the ammonia molecule in CNT.
As MD simulations, besides $q_{\rm N}$ and $q_{\rm H}$, use other parameters of atoms constituting ammonia molecules,
we combine the obtained charges with the OPLS force field geometry.
In particular,
the valence angles and intramolecular distances were controlled
with harmonic terms $k_{\alpha}\cdot(\alpha_{\rm H-N-H} - \alpha_0)^2$ and $k_{r}\cdot(r_{\rm N-H} - r_0)^2$,
where $\alpha_0 = 106.4^{\circ}$ and $r_0 =1.01$~\AA~\cite{Rizzo1999}.
Because of such flexibility,
the dipole moment of ammonia molecules was found to vary within the range of
$1.26 \ldots 1.36$~D (L\"{o}wdin)
or
$1.32 \ldots 1.43$~D (Mulliken)
along the span of simulated temperatures,
see the upper panel of Fig.~\ref{fig03} and Fig.~\ref{fig08} below.
For the sake of comparison, the higher dipole moment of $\mu=1.94$~D for bulk ammonia is reported in Ref.~\cite{Eckl2008}.
Significantly lower dipole moment of water molecules confined in nanotubes ($\mu=1.105$~D) in comparison to the bulk value ($\mu=2.351$~D)
is also known in literature, see, e.g., Refs.~\cite{Mann2003,Ma2017}.
More details about quantum chemistry calculations are given in Appendix~A.

As in the case of confined water in (6,5) CNT~\cite{Ma2017,Druchok2023},
the quantum chemistry predictions are rather diverse and strongly depend on a calculation scheme
(choice of the basis set, the quantum-mechanical method, the population analysis method and the choice of the geometry of the ammonia molecule),
see, e.g., Ref.~\cite{Ertural2019}.
The ambiguity related to the ammonia molecule model, which is required for further MD studies,
can be avoided while performing {\it ab initio} MD simulations~\cite{Marx2009},
however, such calculations are far beyond the focus of the present paper.
In our MD simulations we use the B3LYP data,
i.e., two sets:
$q_{\rm N}=-0.7814$, $q_{\rm H}=-q_{\rm N}/3$ (L\"{o}wdin charges, see Sec.~\ref{sec22})
and
$q_{\rm N}=-0.9582$, $q_{\rm H}=-q_{\rm N}/3$ (Mulliken charges, see Appendix~B).
Both sets yield qualitatively similar results,
although larger Mulliken charge values result in larger dipole-dipole interaction strength
pushing the temperature-driven orientational quasiordering to higher temperatures.

\subsection{Molecular dynamics simulations and results}
\label{sec22}

We use the DL$_{-}$POLY molecular simulation package~\cite{Todorov2006}
to perform a series of MD simulations of ammonia molecules encapsulated inside the CNT with chirality of (6,5) at different temperatures.
The (6,5) CNTs have a diameter of $\approx 7.4$~\AA.
This value denotes the diameter of the circle over the centers of carbon atoms constituting the CNT openings.
The actual interior, available for ammonia molecules, is smaller due to van der Waals sizes of carbons.
Such a small-sized nanopore allows only for a single-file arrangement of molecules.
Furthermore, we consider $N=35$ ammonia molecules inside a CNT of $\approx 170$~{\AA}.
We ran a set of simulations over a temperature range of $10\ldots 240$~K.
The results around the lower temperature boundary should be taken with caution
because of quantum effects which are not accounted for in our MD simulations.

We generated starting configurations consisting of CNTs and ammonia molecules.
The intramolecular geometry and short-range interactions for ammonia were taken from the OPLS force field,
while the charges for nitrogens and hydrogens were optimized within the B3LYP level of approximation (see Sec.~\ref{sec21} and Appendix~A).
The CNT model was taken from Ref.~\cite{Huang2006}, namely, the Lennard-Jones parameters for carbons of nanotube sidewalls.
Usually, the CNT simulations also imply a set of harmonic bonds and angles maintaining the CNT geometry,
however, since ammonia molecules are in the spot of interest, we froze the ideal CNT in vacuum to cut the computational costs.
We mention that CNT sidewall carbons are neutral.
The Lennard-Jones parameters for unlike sites were calculated using the Lorentz-Berthelot mixing rules.
This combination of interaction parameters was successfully utilized in our recent studies~\cite{Druchok2017,Druchok2018,Druchok2019,Druchok2023}
of nanotubes interacting with SPC/E water.
Finally, two additional carbon atoms were placed at the centers of CNT openings, which play a role of obstacles,
to assure that ammonia molecules stay inside the nanotube interior during the whole set of simulations.

The simulation conditions were kept the same for all the runs except the temperature variation.
The temperature was controlled by means of the NVT Nose--Hoover thermostat.
Each simulation utilized the leapfrog integration algorithm with the time step of 0.001~ps,
covering 200~ps of equilibration and then 600~ps of production runs.
Smooth particle mesh Ewald technique was used to calculate the electrostatic terms,
while the short-range interactions were calculated with the cut-off distance of 15~{\AA}.

We also performed MD simulations for $N=35$ water molecules in the (6,5) CNT of $\approx170$~{\AA}
(the results are collected in Appendix~C)
for comparison with the ones for the ammonia case.

\begin{figure}
\begin{center}
\includegraphics[clip=true,width=0.9\columnwidth]{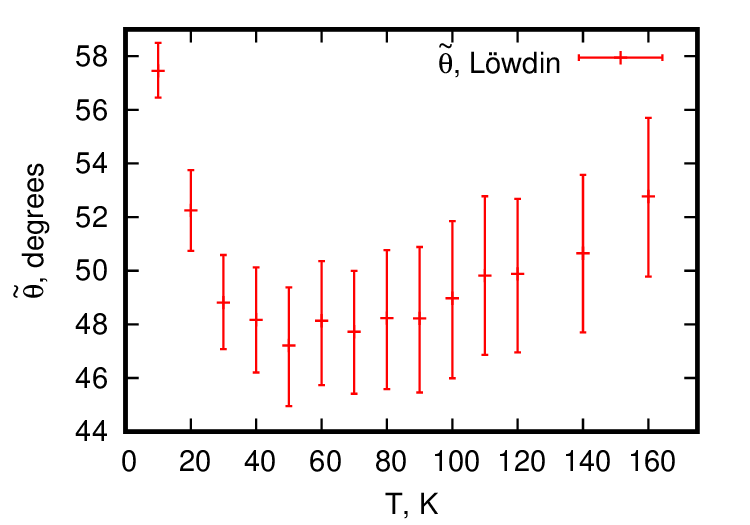}
\includegraphics[clip=true,width=0.9\columnwidth]{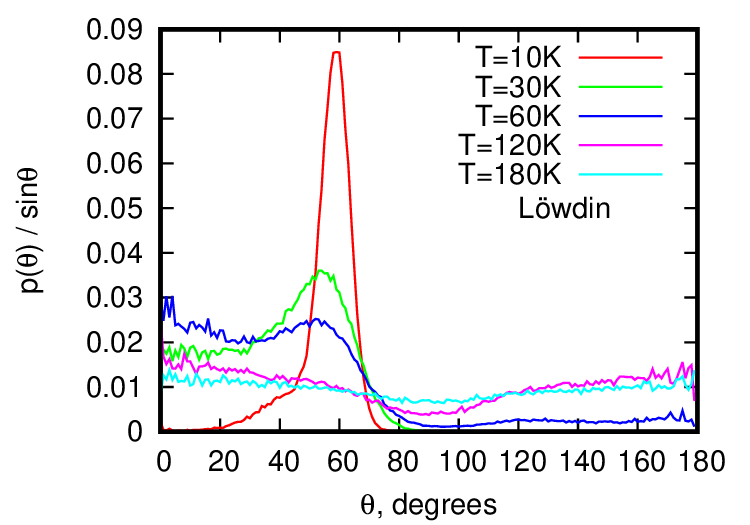}
\caption{Angles between CNT axis and dipole moments of ammonia molecules.
Top panel: Temperature dependence of mean angles averaged over time and ensemble $\tilde{\theta}$.
Bottom panel: $p(\theta)/\sin\theta$ for multiple temperatures;
$p(\theta)/\sin\theta$ has maximum at $\theta\approx57^\circ$ for $T=10$~K.}
\label{fig02}
\end{center}
\end{figure}

\begin{figure}
\begin{center}
\includegraphics[clip=true,width=0.9\columnwidth]{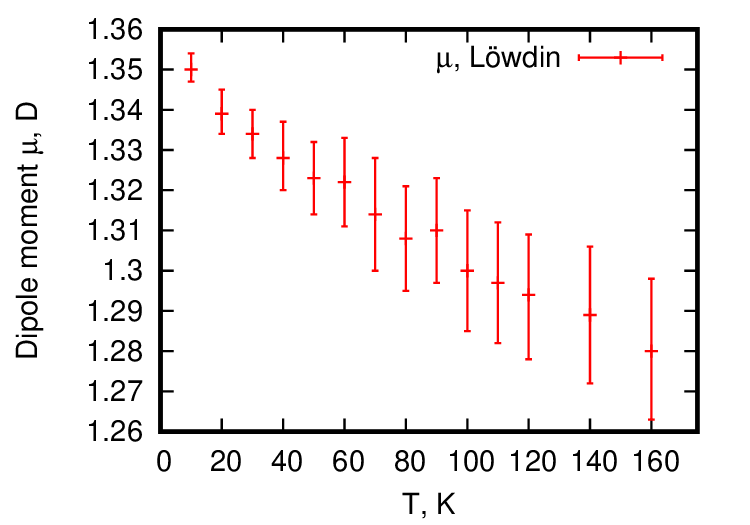}
\includegraphics[clip=true,width=0.9\columnwidth]{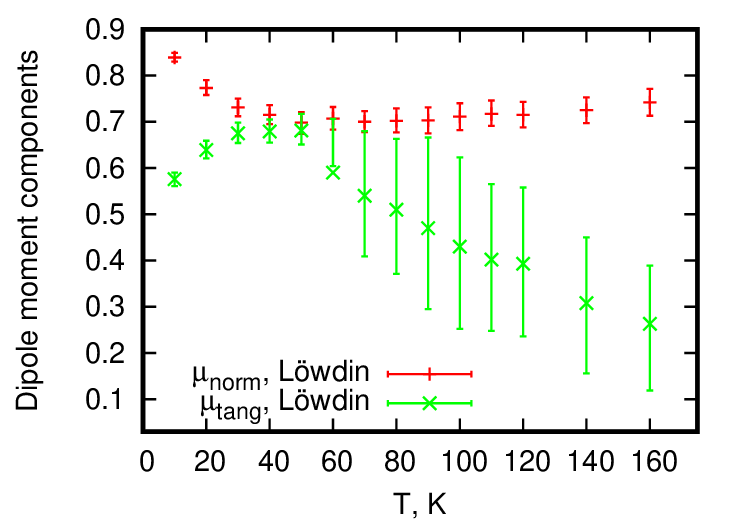}
\caption{Top panel:
Temperature dependence of ammonia dipole moment $\mu$.
Bottom panel:
Mean magnitudes of
the perpendicular component of dipole moment for ammonia molecule (red)
and
the tangential component of total dipole moment of ammonia chain per ammonia molecule (green);
both components are additionally normalized by $\mu$.}
\label{fig03}
\end{center}
\end{figure}

\begin{figure}
\begin{center}
\includegraphics[clip=true,width=0.9\columnwidth]{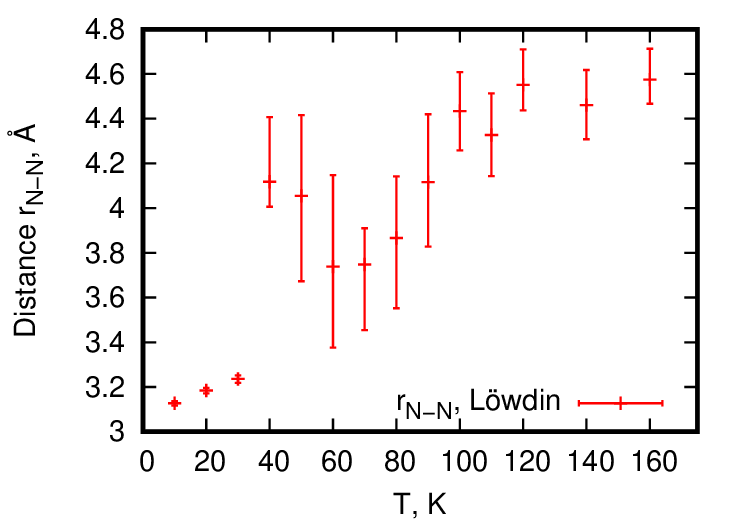}
\caption{Temperature dependence of mean distances between nearest nitrogen atoms within ammonia chain.}
\label{fig04}
\end{center}
\end{figure}

Our MD simulation results for the ammonia molecules in the (6,5) CNT are reported in Figs.~\ref{fig02}, \ref{fig03}, and \ref{fig04}.
These results
i) illustrate the emergence of the intermediate ordered quasiphase
and
ii) provide an input for the lattice model to be considered in Sec.~\ref{sec3}.

Figure~\ref{fig02} illustrates temperature dependence of orientations of confined ammonia molecules.
Due to an arbitrary choice of the CNT axis direction,
the orientational order within the ammonia chain can be characterized in two manners -- ``individual'' and ``group'' ones.
In general, the angle $\theta$ between the dipole moment of ammonia molecule and the CNT axis can take values from the range of $0\ldots180^\circ$.
By adopting the same axis direction for all dipoles we refer to this as the group orientation.
The definition of $\tilde{\theta}$ as $\min(\theta, 180^\circ - \theta)$ has a meaning of an individual angle.
Both definitions make sense,
as $\tilde{\theta}$ is more focused on independent orientation of a dipole and is useful for our lattice model,
while $\theta$ distribution brings information about mutual orientation of dipoles along the chain.
Both definitions, in their own way, can serve as indicators of temperature-induced rearrangement of dipoles.
In particular, the top panel of Fig.~\ref{fig02} demonstrates evolution of ``individual'' angles $\tilde{\theta}$.
Here and below in plots with MD results,
besides the mean values, we also show errorbars to indicate 25-th and 75-th percentiles of values collected during the production runs of simulations.
The mean value for $\tilde{\theta}$ at $T=10$~K is about 57$^\circ$ and decreases down to $47^\circ$ at $T=50$~K,
then increases with the temperature raise.
One can also see that the errorbars start to span over larger intervals with the temperature growth.
The bottom panel of Fig.~\ref{fig02} demonstrates the probability distribution density of a ``group'' angles $\theta$
normalized by $\sin\theta$.
Such a denominator is needed to account for different number of redundant states for different $\theta$.
One can see that at low temperature ($T=10$~K) the distribution $p(\theta)/\sin\theta$ is focused in vicinity of $\theta\approx 57^\circ$,
which corresponds to the value of $\tilde{\theta}\approx 57^\circ$ spotted above.
``Individual'' and ``group'' angles coincide, thus, the chain demonstrates a uniform orientation of dipoles.
With the temperature rise, the distributions $p(\theta)/\sin\theta$ at first spread to smaller angles,
indicating the intermediate quasiorder ($T=30$ and 60~K),
then, become broader with appearance of angles about 90$^\circ$,
indicating that molecules can flip their direction ($T = 120$ and 180~K).
Such a behavior of $p(\theta)/\sin\theta$ may also yield a rough estimate for the temperature interval $T_1 < T_2$
where the intermediate quasiphase with dipole moments parallel to the CNT axis exists.

Top panel of Fig.~\ref{fig03} shows temperature dependence of the ammonia molecules dipole moment.
As can be seen in the top panel of Fig.~\ref{fig03},
the higher the temperature the higher variation of dipole values around means,
speaking in favor of increasing fluctuations in the system.

In the bottom panel of Fig.~\ref{fig03} we show
the normal component of dipole moment of individual ammonia molecules $\mu_{\rm norm} = \overline{\vert\mu_{\perp}\vert}/\mu$
and
the total dipole moment of ammonia chain tangential to the CNT axis $\mu_{\rm tang} = \overline{\mu^{\rm tot}_{\parallel}}/(N \mu)$;
here $\overline{(\ldots)}$ denotes the mean value of $(\ldots)$.
Clearly,
$\mu_{\rm tang}$ and $\mu_{\rm norm}$ reach maximum and minimum, correspondingly, in vicinity of $T=40\ldots50$~K
as expected for the intermediate quasiphase with dipole moments aligned along the CNT axis.

X.~Ma {\it et al.}~\cite{Ma2017} observed similar temperature profiles for confined water
and pointed out three types of water arrangement:
1) hydrogen-bonding over the whole chain, when dipole moments of water molecules are tilted by 31$^\circ$ to the CNT axis
(quasiphase 1),
2) dipole moments tend to align along the CNT axis in one direction
(quasiphase 2),
and
3) collective arrangement is completely destroyed
(quasiphase 3).
On the basis of similar results reported in Fig.~\ref{fig03},
we may assume that the quasiphase 2 is achieved for ammonia in vicinity of $T=40\ldots50$~K,
while the quasiphases 1 and 3 are located at lower and higher temperatures, respectively.

In addition to MD analysis of the dipole moment magnitudes and orientations,
mean distances between nearest ammonia molecules, as they follow from MD study, are also an important reference.
We monitored nitrogen-nitrogen distances during the simulation and then averaged them; such an average is reported in Fig.~\ref{fig04}.
Interesting to note, that both mean N-N distance and its variation increase once the system passes the temperature range of quasiphase 2.

We end up the section with a remark about ammonia molecule model based on the Mulliken charges.
The Mulliken case demonstrates a similar behavior as in Figs.~\ref{fig02}, \ref{fig03}, and \ref{fig04}
(cf. Figs.~\ref{fig07}, \ref{fig08}, and \ref{fig09} in Appendix~B),
i.e.,
predicts three quasiphases and two quasiphase transitions between them,
however, the temperature range for the intermediate quasiphase 2 is now $T=90\ldots100$~K.

\section{Lattice model}
\label{sec3}

\subsection{Lattice model formulation}
\label{sec31}

We pass to statistical mechanics arguments
for explaining behavior of ammonia molecules forming a single-file chain in CNT and emergence of the intermediate quasiphase~\cite{Druchok2023}.
We bear in mind that the system at hand is one-dimensional and consists of finite (not very large) number of molecules $N$.
We have to account for
short-range nearest-neighbor interactions leading to hydrogen-bonded chains
and
long-range dipole-dipole interactions
as well as
rotations with limitations imposed by the geometry of CNT having
a very small diameter that can only be filled with one ammonia molecule after the other \cite{Yang2020}.
Then, an interplay of these ingredients produces a finite range of temperatures in which the states,
in which ammonia molecule dipole moments are directed along the CNT axis,
dominate in the state space~\cite{Druchok2023}.

More specifically,
we consider a lattice model with $3$ states at each lattice site subjected to certain restrictions
(hydrogen-bonded chain consists at least of two molecules)
which effectively decrease the number of states per site.
Thus,
we deal with $N$ rigid bodies each with the moment of inertia $I$,
which carry coplanar dipoles $\vec{\mu}_j$, $j=1,\ldots,N$;
they are rendered on a single straight line so that the distance between the neighboring sites $j$ and $j+1$ is $a_{j,j+1}$.
Each site $j$ may be in one of the following 3 states $\xi_j$:
\begin{itemize}
\item
The state $\xi_j=1$,
when $\mu_{\parallel,j}=\mu \cos \alpha_1$, $\vert\mu_{\perp,j}\vert=\mu \sin \alpha_1$
and the extension of the occupied site is $a_1$.
The site being in such a state belongs to a hydrogen-bonded chain
and must have at least one neighboring site in the same state $\xi=1$.
Furthermore, the set of $\mu_{\perp,j}$ for the hydrogen-bonded chain forms a staggered pattern.
\item
The state $\xi_j=2$,
when $\mu_{\parallel,j}=\mu \cos \alpha_2$, $\vert\mu_{\perp,j}\vert=\mu \sin \alpha_2$, $0^{\circ}\le\alpha_2<\alpha_1$,
and the extension of the occupied site is $a_2=a_1(1+\varepsilon_2)>a_1$.
\item
The state $\xi_j=3$,
$\mu_{\parallel,j}=\mu \cos \alpha_3$, $\vert\mu_{\perp,j}\vert=\mu \sin \alpha_3$, $\alpha_1<\alpha_3\le 90^{\circ}$,
and the extension of the occupied site is $a_3=a_1(1+\varepsilon_3)>a_2$.
This state represents a completely independent ammonia molecule with a random orientation of $\vec{\mu}_j$.
\end{itemize}
The introduced rules reduce the number of states $W_N$ for the lattice of $N$ sites, which is now $W_{N}\approx 2.52^N<3^N$~\cite{Druchok2023}
(i.e., the lattice model has $\approx 2.52<3$ states per each site).

From the mathematical point of view,
$W_N$ is equal to the number of words of length $N$ over the three-letter alphabet $\{1,2,3\}$
without words having isolated 1's~\cite{Birmajer2016,Sloane1964}.
Moreover,
$W_N$ is defined recursively according to the  recurrence relation
$W_N=3W_{N-1}-2W_{N-2}+2W_{N-3}$, $W_0=1$, $W_1=2$, $W_2=5$
and can be obtained from the generating function $g(z)=\sum_{N=0}^{\infty}z^NW_N$,
which is given by $g(z)=(1-z+z^2)/(1-3z+2z^2-2z^3)$,
see Ref.~\cite{Sloane1964}.
Furthermore~\cite{Hall1967}, inserting $W_N\propto a^N$ for $N\to \infty$ into the recurrence relation,
one gets a cubic equation for $a$,
$a^3-3a^2+2a-2=0$,
the largest root of which
$(27+3\sqrt{78})^{1/3}/3+(27+3\sqrt{78})^{-1/3}+1\approx 2.521\,379\,706\,8$
settles $W_N \approx 2.52^N$ as $N\to\infty$.

Now, the thermodynamic average is defined as follows:
\begin{eqnarray}
\label{01}
\langle(\ldots)\rangle
=\frac{1}{Z}
{\sum_{\xi_1\ldots\xi_N}}^{\prime}\sum_{\rm rot}
\left(\exp\left[-\frac{E(\xi_1 \ldots \xi_N)}{k_{\rm B}T}\right](\ldots)\right),
\nonumber\\
Z={\sum_{\xi_1\ldots\xi_N}}^{\prime}\sum_{\rm rot}\exp\left[-\frac{E(\xi_1\ldots\xi_N)}{k_{\rm B}T}\right].
\end{eqnarray}
Here
the prime near the first sum indicates the discussed above restriction on the set of values $\xi_1 \ldots \xi_N$ (no words with isolated 1's)
and
the second sum denotes the summation over rotational degrees of freedom for given (allowed) set $\xi_1 \ldots \xi_N$.
Moreover,
$E(\xi_1 \ldots \xi_N)$ stands for the sum of the rotation energy and the interaction energy
which contribute to
the rotation part $K$
and
the interaction part $Q$
of the partition function $Z=Z(T,N)$.

We take into account the short-range nearest-neighbor interactions
by treating the molecules at sites $j$ and $j+1$ with $a_{j,j+1}=a_1$ as linked (i.e., rigidly connected) through a  hydrogen bond.
More generally,
we have to assume in addition that the energy of the hydrogen-bonded chain also decreases because of the bonding, see below.
The long-range interactions between all molecules $U_{\xi_1\ldots\xi_N}$
is the sum over all $N(N-1)/2$ pairs of the dipole-dipole interaction $u_{ij}$, $i<j$, $i=1,\ldots,N-1$, $j=2,\ldots,N$.
Moreover,
\begin{eqnarray}
\label{02}
u_{ij}=k\frac{\mu_{\perp,i}\mu_{\perp,j}-2\mu_{\parallel,i}\mu_{\parallel,j}}{a^3_{ij}},
\end{eqnarray}
$k=1/(4\pi\epsilon_0)$, $\epsilon_0$ is the vacuum permittivity (SI units),
if the both sites $i$ and $j$ belong to the same hydrogen-bonded chain.
However,
\begin{eqnarray}
\label{03}
u_{ij}=k\frac{-2\mu_{\parallel,i}\mu_{\parallel,j}}{a^3_{ij}},
\end{eqnarray}
if the sites $i$ and $j$ belong to different hydrogen-bonded chains or at least one of these sites is in the state 2 or 3.
In other words,
the $\mu_{\perp}$ on-site components contribute to the intersite interaction $u_{ij}$ only if the both sites rotate as a whole,
but do not contribute to the intersite interaction if they rotate independently.
In contrast,
the $\mu_{\parallel}$ on-site components always contribute to the intersite interaction $u_{ij}$.

Limited (because of CNT geometry) rotations are accounted as follows.
A hydrogen-bonded chain consisting of $n$ molecules has the moment of inertia $nI$ and rotates along one axis only,
which coincides with the nanotube axis.
Its energy is given by $E_m=\hbar^2 m^2/(2nI)$ with $m=0,\pm 1,\pm 2,\ldots$
and hence each energy level $E_m$ except the one with $m=0$ is two-fold degenerate~\cite{Galitski1981}.
The rotational partition function of the hydrogen-bonded chain reads:
\begin{eqnarray}
\label{04}
K_{n}^{(1)}=\sum_{m=-\infty}^\infty\exp\left(-\frac{m^2}{n\tau}\right),
\end{eqnarray}
$\tau=T/T_{\rm rot}$,
$k_{\rm B}T_{\rm rot}=\hbar^2/(2I)$.
Furthermore,
we assume that a site being in the state 2 corresponds to a molecule which rotates similarly to the hydrogen-bonded chain,
i.e., contributes $K_{1}^{(1)}$ to the rotation part of the partition function.
Moreover,
a site being in the state 3 corresponds to a molecule which rotates along three axes;
its energy is given by $E_J=\hbar^2 J(J+1)/(2I)$ with $J=0,1,2,\ldots$
and the degeneracy of the energy level $E_J$ is $(2J+1)^2$.
The partition function of such a rotor reads:
\begin{eqnarray}
\label{05}
K_{1}^{(3)}=\sum_{J=0}^\infty\left(\left(2J+1\right)^2\exp\left[-\frac{J(J+1)}{\tau}\right]\right).
\end{eqnarray}

Having the partition function $Z$ (\ref{01}),
we immediately get the Helmholtz free energy $F=-k_{\rm B}T\ln Z$ and hence
the entropy $S=-\partial F/\partial T$,
the internal energy $E=F+TS$,
and
the specific heat $C=T\partial S/\partial T$.
According to Eq.~(\ref{01}) we also get the average dipole moment (per site) or, more precisely, the following quantities:
\begin{eqnarray}
\label{06}
\mu_{\parallel}=\frac{1}{N}\sum_{j=1}^N\langle\mu_{\parallel,j}\rangle,
\;\;\;
\vert\mu_{\perp}\vert=\frac{1}{N}\sum_{j=1}^N\langle\vert\mu_{\perp,j}\vert\rangle.
\end{eqnarray}
Obviously,
$\mu_{\rm tang}=\mu_{\parallel}/\mu$
and
$\mu_{\rm norm}=\vert\mu_{\perp}\vert/\mu$,
see Sec.~\ref{sec22} and Fig.~\ref{fig03}.
Finally,
we can calculate
the average length of the chain $L$
and
the coefficient of linear thermal expansion $\alpha_L=(1/L)(\partial L/\partial T)$:
\begin{eqnarray}
\label{07}
L=\sum_{j=1}^{N-1}\langle a_{j,j+1}\rangle,
\;\;\;
\alpha_L=\frac{1}{L}\frac{{\rm d} L}{{\rm d} T}.
\end{eqnarray}
The length of the chain per site $L/N$ might be related to the mean distance between the nearest nitrogen atoms,
see Fig.~\ref{fig04}.

There are only a few parameters which are used as the input for the lattice model described above.
We begin with the energy scale which is determined by $T_{\rm rot}$.
Using the values $I_A=I_B=2.8\times 10^{-47}$ or $I_C=4.4\times 10^{-47}$ in units of SI~\cite{Badger1929,David1996}
we obtain for $T_{\rm rot}$ the values about $14$ or $9$~K.
In our calculations we set $T_{\rm rot}=10$~K.
The energy scales which is determined by $T_{\rm dip}\propto(k\mu^2/a^3)/k_{\rm B}$
depends on the values of the dipole moment $\mu$ and the characteristic length $a$.
The quantum-chemical computations and MD simulations illustrated in Sec.~\ref{sec2} allow us to choose these parameters.
Using data for $\mu=1.35\ldots1.28$~D and $a=3.1\ldots4.6$~{\AA} from Figs.~\ref{fig03} and \ref{fig04} (L\"{o}wdin charges),
we may set $R=0.07$, that is, $T_{\rm dip}=T_{\rm rot}/R\approx143$~K.
Note that for the Mulliken charges, Figs.~\ref{fig08} and \ref{fig09}, one has to decrease $R$ slightly.
Next, we fix the angles as follows:
$\alpha_1=56^\circ$, $\alpha_2=20^\circ$, $\alpha_3=80^\circ$,
cf. Fig.~\ref{fig02}.
Finally, the length scale is determined by $a_1$, $a_2$, and $a_3$:
It might be reasonable to set
$a_1=3.10$~{\AA},
$\varepsilon=0.25$,
$a_2=(1+\varepsilon)a_1\approx3.88$~\AA,
and
$a_3=(1+2\varepsilon)a_1=4.65$~{\AA}
in view of the MD results reported in Fig.~\ref{fig04}.
The values $0^\circ\le\alpha_2<\alpha_1<\alpha_3\le90^\circ$ and $a_1<a_2<a_3$ are chosen to be in line with whole temperature dependencies
shown in the bottom panel of Fig.~\ref{fig03} and in Fig.~\ref{fig04}.

Importantly, the value of $\varepsilon$ must exceed a certain threshold value
in order to have as the ground state the hydrogen-bonded chain $111\ldots$ rather than the state $22\ldots2$.
More realistic description mentioned above would imply including an energy of the hydrogen-bonded chain
which diminishes the energy of the state $111\ldots$ .
In the present study, we do not take into account the bonding energy which may modify quantitatively the lattice-model outcome.

We emphasize here that our aim is not to reproduce MD simulations by the lattice model analysis,
but only to demonstrate a reasonable agreement between the conclusions yielded by these approaches,
i.e., MD simulations and statistical mechanics calculations.

\subsection{Lattice model results}
\label{sec32}

\begin{figure}
\begin{center}
\includegraphics[clip=true,width=0.9\columnwidth]{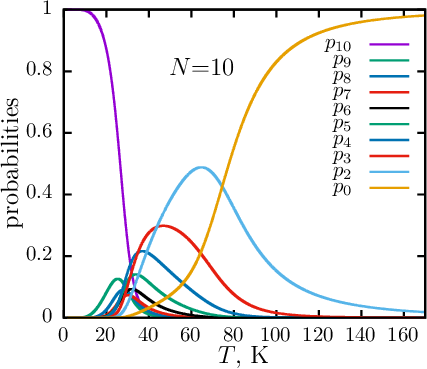}\\
\includegraphics[clip=true,width=0.9\columnwidth]{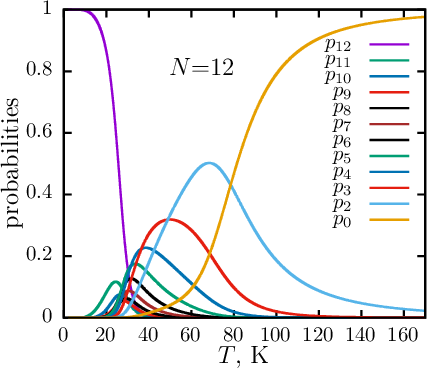}
\caption{Probabilities of various configurations versus temperature for the chains of $N=10$ (top) and $N=12$ (bottom) sites.
There is a temperature region $50\ldots75$~K where the probability $p_2$ (blue curves) dominates.}
\label{fig05}
\end{center}
\end{figure}

\begin{figure}
\begin{center}
\includegraphics[clip=true,width=0.9\columnwidth]{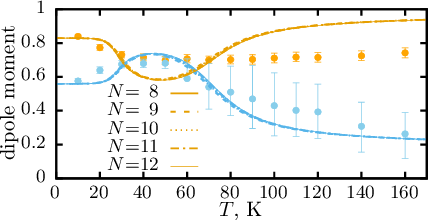}\\
\includegraphics[clip=true,width=0.9\columnwidth]{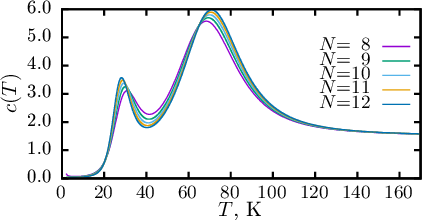}\\
\includegraphics[clip=true,width=0.9\columnwidth]{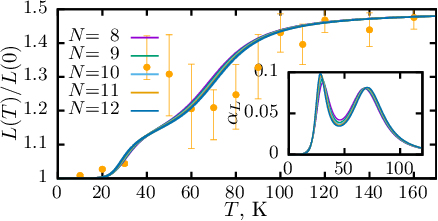}\\
\includegraphics[clip=true,width=0.9\columnwidth]{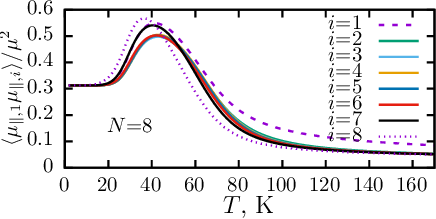}
\caption{Temperature dependencies of
(from top to bottom)
$\mu_{\parallel}/\mu$ (blue) and $\vert\mu_{\perp}\vert/\mu$ (orange),
see Eq.~(6),
specific heat $c(T)=C/(k_{\rm B}N)$,
average length (coefficient of linear thermal expansion is shown the inset), see Eq.~(7),
and
correlators $\langle \mu_{\parallel,1} \mu_{\parallel,i}\rangle/\mu^2$, $i=1,\ldots,N$
for the chains of $N{=}8,\ldots,12$ sites.
MD simulations
($N=35$ ammonia molecules in the CNT of length $\approx170$~{\AA}, filled circles) are shown in the panels
with $\mu_{\parallel}/\mu$, $\vert\mu_{\perp}\vert/\mu$ and the average length [we set $L(0)/(N-1)=3.10$~{\AA}].}
\label{fig06}
\end{center}
\end{figure}

We perform all lattice-model calculations described in previous subsection for $N=8,9,10,11,12$
using the Maple software package implemented on a personal computer.
Our findings are reported in Figs.~\ref{fig05} and \ref{fig06} and are discussed below.

To illustrate how the intermediate phase shows up,
we single out in the partition function $Z$ (\ref{01}) the contributions of hydrogen-bonded chains of various lengths.
Namely,
of the length $N$, ${\cal {Z}}_N=K_N^{(1)}{\cal {Q}}_{N}$,
of the length $N-1$, ${\cal {Z}}_{N-1}=K_{N-1}^{(1)}{\cal {Q}}_{N-1}$,
\ldots,
of the length 2, ${\cal {Z}}_{2}=K_{2}^{(1)}{\cal {Q}}_{2}$,
and the contributions without the hydrogen-bonded chains ${\cal{Z}}_0$,
i.e.,
$Z={\cal {Z}}_N+{\cal {Z}}_{N-1}+\ldots+{\cal {Z}}_2+{\cal {Z}}_0$,
where $K_n^{(1)}$ is given in Eq.~(\ref{04}) and ${\cal {Q}}_{n}$ denotes the remaining part of ${\cal {Z}}_n$, $n=N,\ldots,2$,
see Ref.~\cite{Druchok2023}.
Moreover, we introduce the probabilities
$p_N={\cal {Z}}_N/Z$, $p_{N-1}={\cal {Z}}_{N-1}/Z$, \ldots, $p_2={\cal {Z}}_2/Z$, and $p_0={\cal {Z}}_0/Z$;
obviously, $p_{N}+p_{N-1}+\ldots+p_2+p_0=1$.

The temperature-dependent probabilities $p_N$, $p_{N-1}$, \ldots, $p_2$, and $p_0$
control the role of the configurations with different length of hydrogen-bonded chains in thermodynamics.
In the zero-temperature limit $T\to 0$,
when the lowest-energy ground state dominates,
$Z\to {\cal{Z}}_N$ and $p_N\to1$.
In the high-temperature limit $T\to\infty$,
when the dipole-dipole interactions become irrelevant
$Z\to {\cal{Z}}_0 \to (K_1^{(3)})^N$ and $p_0\to 1$.

Temperature dependencies of $p_N$, $p_{N-1}$, \ldots $p_2$, and $p_0$ for $N=10$ and $N=12$ are shown in Fig.~\ref{fig05}.
As can be seen from this figure,
there is a wide temperature range of $55\ldots75$~K where the largest probability $p_2$ exceeds 40\% for both values of $N$.
More detailed analysis of ${\cal Q}_2$ shows
that the main contribution to $p_2$ comes from the subset of configurations
in which the remaining $N-2$ molecules are in the state 2~\cite{Druchok2023}.

In summary,
the considered probabilities $p_N,\ldots,p_0$ for $N=8,\,9,\,10,\,11,\,12$
provide evidence that the states,
which contain ammonia molecules in the on-site states 1 and 2,
are the most relevant ones in the temperature range $50\ldots75$~K
and result in the emergence of an intermediate quasiphase.
Since the ammonia molecules in the state 2 and in the short hydrogen-bonded chains contribute to $\mu_{\parallel}$ but not to $\mu_{\perp}$,
the tangential/normal component of total dipole moment should increase/decrease in this temperature interval.
Further evidence for that follows from analysis of temperature dependencies of observable quantities to be discussed below.

The intermediate quasiphase is stable in a rather wide temperature region $T_1\ldots T_2$.
An estimate for the temperatures of quasiphase transitions $T_1<T_2$ can be drawn by equating the corresponding probabilities,
that is,
$p_3(T_1)=p_2(T_1)$
and
$p_2(T_2)=p_0(T_2)$.
For the chosen set of parameters we get $T_1\approx39,\,41,\,44,\,46,\,49$~K and $T_2\approx69,\,72,\,75,\,77,\,79$~K as $N=8,\,9,\,10,\,11,\,12$.

In Fig.~\ref{fig06} we show temperature dependencies for various quantities of interest for the lattice model of $N=8,9,10,11,12$ sites.
The introduced model predicts
an increase of $\mu_{\parallel}$ (\ref{06}) and decrease of $\vert \mu_{\perp}\vert$ (\ref{06}) in the temperature range $30\ldots80$~K
in agreement with MD simulations,
see blue and orange curves in the top panel of Fig.~\ref{fig06}.
The specific heat per site in the interval $30\ldots70$~K has minimum with the minimal value which noticeably depends on $N$.
At high temperatures the specific heat approaches $3k_{\rm B}/2$ as it should for independent three-axes rotators,
see Eq.~(\ref{05}).
The average length $L$ increases with the temperature growth, however, differently at different temperatures.
This can be also seen from the temperature dependence of $\alpha_L$ from Eq.~(\ref{07}) shown in the inset.
We plot also the results of MD simulations shown previously in Fig.~\ref{fig04} after assuming $L(0)/(N-1)=3.10$~{\AA}.
Again both results, lines and filled circles, agree qualitatively.

In the lowest panel of Fig.~\ref{fig06} we report the lattice-model predictions for the dipole correlators
$\langle \mu_{\parallel,1} \mu_{\parallel,i}\rangle$, $i=1,\ldots,8$ ($N=8$),
which also indicate a correlated state of ammonia molecules in CNT at low temperatures.
From this figure we conclude that all correlators practically coincide,
i.e., are independent on distance between the sites,
up to 15~K
having the value about $\cos^256^\circ$ 
(quasiphase 1 when molecules form a hydrogen-bonded chain),
and with further temperature growth up to 55~K the correlators with $i=2,\ldots,6$ remain almost indistinguishable
passing the maximal value about 0.5
(quasiphase 2).
Only for higher temperatures,
the dipole correlations decrease with increase of the distance between sites
and gradually fall down to $\cos^280^\circ$ 
(quasiphase 3).

In the end of subsection we remark that the estimates of the temperature interval for the intermediate quasiphase,
as they follow from inspection of various quantities,
are slightly different and depend on the quantity under analysis.
This is yet another indication that we face a gradual emergence of the intermediate quasiphase but not a strict phase transition.

\section{Discussion and summary}
\label{sec4}

In conclusion,
motivated by the experiments reported in Ref.~\cite{Ma2017} and their theoretical support \cite{Ma2017,Druchok2023},
which demonstrated a temperature-driven water dipole quasiordering in one dimension,
we address a question whether such phenomenon can be observed for other molecules and focus on the ammonia molecules NH$_3$.
Our theoretical study,
which includes quantum chemistry calculations for the model of ammonia molecule and MD simulations for ammonia molecules encapsulated in (6,5) CNT
as well as statistical mechanics arguments on the basis of a lattice model,
gives an affirmative answer to this question:
Ammonia dipoles show quasiorder below 100~K becoming parallel to the CNT axis.
Even though details of quasiordering depend on the input characteristics of the ammonia molecule inside CNT
(L\"owdin or Mulliken charges)
the quasiordering cannot be questioned.

Both the water molecules and the ammonia molecules show similar behavior.
Namely,
atomic charges in CNT are reduced and rotations within CNT becomes restricted,
the temperature dependence of the probability distribution density of the angle $\theta$ between the molecule dipole moment and the CNT axis indicate three phases and two quasiphase transitions between them,
an interplay of the interaction contribution and the entropy contribution
(rotations within restricted one-dimensional geometry)
produce a highly ordered structures in a temperature range $T_1\ldots T_2$.
In this temperature range, as it follows from the lattice-model analysis,
the states with dipole moments oriented along the CNT axis dominate the partition function
that can be seen in the temperature dependencies of
$\mu_{\parallel}$, $\vert\mu_{\perp}\vert$,
the specific heat
or
dipole correlators.
Clearly, we deal with a finite number of sites one-dimensional lattice model and any true phase transitions cannot be expected,
however, a gradual replacement of one quasiphase by another is possible and several calculated quantities do indicate this.
The quasiordering at intermediate temperatures is a classical phenomenon:
It occurs at temperatures at which the thermal de Broglie wavelength of ammonia molecule is much smaller than, e.g., intermolecular distances.

Most evident difference between the water molecules and the ammonia molecules
is the angle of molecule dipoles relative to the nanotube axis in the hydrogen-bonded quasiphase,
which is about two times larger for ammonia case in comparison the water case
(recall, $\alpha_1=31^\circ$ for water \cite{Druchok2023} versus  $\alpha_1=56^\circ$ for ammonia).

The most interesting question which naturally shows up, is whether it is possible
to encapsulate ammonia molecules inside a narrow CNT
and then examine extensively ammonia-filled and empty CNTs for comparison.
Unfortunately, we are not aware about this.
We hope that the presented theoretical study may be of use for corresponding experiments in the future.

\section*{Acknowledgments}

The authors thank Danylo Dobushovskyi for bringing Refs.~\cite{Sloane1964,Birmajer2016} to their attention.

\section*{Author declarations}

{\bf Conflict of interest:}

The authors have no conflicts to disclose.

{\bf Author contributions:}

O.~D. conceived the study;
M.~D. performed molecular dynamics simulations;
V.~K. performed quantum chemical calculations;
T.~K. performed calculations for the lattice model.
All authors discussed the results and commented on the manuscript.

\section*{Data availability}

The data that support the findings of this study are available within the article.

\appendix
\section*{Appendix~A: Quantum chemistry calculation details}
\renewcommand{\theequation}{A.\arabic{equation}}
\setcounter{equation}{0}

\begin{table}
\caption{Quantum chemistry calculations for the ammonia molecules restricted to one dimension (without CNT),
see Sec.~\ref{sec21}.
Semi-empirical methods AM1 and PM6.}
\vspace{2mm}
\begin{center}
\begin{tabular}{|c|c|c|}
\hline
                                  & AM1       & PM6
\\
\hline
$q_{\rm N}$ ($e$)                 & $-0.3965$ & $-0.6127$
\\
\hline
$q_{\rm H}$ ($e$)                 &  0.1322   &  0.2042
\tabularnewline
\hline
$\alpha_{\rm H-N-H}$ ($^{\circ}$) &  107.8    & 105.7
\tabularnewline
\hline
$r_{\rm N-N}$ (\AA)               &  2.95     & 2.97
\tabularnewline
\hline
$r_{\rm N-H}$ (\AA)               & 1.00      &  1.01
\tabularnewline
\hline
\end{tabular}
\end{center}
\label{tab1}
\end{table}

\begin{table}
\caption{Quantum chemistry calculations for the ammonia molecules inside the CNT,
see Sec.~\ref{sec21}.
Semi-empirical methods AM1 and PM6.}
\vspace{2mm}
\begin{center}
\begin{tabular}{|c|c|c|}
\hline
                                   & AM1        & PM6
\tabularnewline
\hline
$q_{\rm N}$ ($e$)                  & $-0.3913$  & $-0.6281$
\\
\hline
$q_{\rm H}$ ($e$)                  & $0.1289$   &  $0.2100$
\tabularnewline
\hline
$\alpha_{\rm H-N-H}$ ($^{\circ}$)  & $107.6$    & $105.1$
\tabularnewline
\hline
$r_{\rm N-N}$ (\AA)                & $3.05$     & $2.94$
\tabularnewline
\hline
$r_{\rm N-H}$ (\AA)                & $1.00$     & $1.01$
\tabularnewline
\hline
\end{tabular}
\end{center}
\label{tab2}
\end{table}

\begin{table}
\caption{Quantum chemistry calculations for ammonia molecules restricted to one dimension (without CNT),
see Sec.~\ref{sec21}.
Hartree-Fock method results with the STO-2G and Huzinaga MINI basis sets.
A hydrogen-bonded chain implies,
first,
a forming bond hydrogen charge $q_{\rm H}$ (the first number before slash in the third row)
and
two dangling hydrogens having charge $q_{\rm H}$ (the second number after slash in the third row)
and,
second,
a shorter and a longer ${\rm N}$-${\rm H}$ bonds with $r_{\rm N-H}$ given by the first and the third numbers in the last row, respectively,
as well as the dangling hydrogens with $r_{\rm N-H}$ given by the second number in the last row.
The partial charges $q_{\rm N}$ and $q_{\rm H}$ are determined
from the L\"{o}wdin population analysis (the upper rows in the second and third rows)
or
from the Mulliken population analysis (the lower rows in the second and third rows).}
\vspace{2mm}
\begin{center}
\begin{tabular}{|c|c|c|}
\hline
                   & STO-2G                                                    &  MINI
\tabularnewline
\hline
$q_{\rm N}$ ($e$)  & $\begin{array}{c}-0.2352\\-0.3735\end{array}$             & $\begin{array}{c}-0.4272\\-0.6150\end{array}$
\\
\hline
$q_{\rm H}$ ($e$)  & $\begin{array}{c}0.1249/0.0553\\0.1821/0.0958\end{array}$ & $\begin{array}{c}0.1662/0.1317\\0.2350/0.1911\end{array}$
\tabularnewline
\hline
$\alpha_{\rm H-N-H}$ ($^{\circ}$) & 103.5/101.9                                & 108.3/107.8
\tabularnewline
\hline
$r_{\rm N-N}$ (\AA) & 2.83                                                     & 3.29
\tabularnewline
\hline
$r_{\rm N-H}$ (\AA) & 1.06/1.04                                                & 1.03/1.02
\tabularnewline
\hline
\end{tabular}
\end{center}
\label{tab3}
\end{table}

\begin{table}
\caption{Quantum chemistry calculations for the ammonia molecule restricted to one dimension (without CNT),
see Sec.~\ref{sec21}.
Density-functional-theory method (B3LYP) results,
see explanations in the title of Table~\ref{tab3}.}
\vspace{2mm}
\begin{center}
\begin{tabular}{|c|c|}
\hline
                    & B3LYP
\tabularnewline
\hline
$q_{\rm N}$ ($e$)   & $\begin{array}{c}-0.7814\\-0.9582\end{array}$
\\
\hline
$q_{\rm H}$ ($e$)   & $\begin{array}{c}0.2704/0.2540\\0.3761/0.2918\end{array}$
\tabularnewline
\hline
$\alpha_{\rm H-N-H}$ ($^{\circ}$)  & 105.5/104.9
\tabularnewline
\hline
$r_{\rm N-N}$ (\AA)   & 3.18
\tabularnewline
\hline
$r_{\rm N-H}$ (\AA)   & $1.03/1.02$
\tabularnewline
\hline
\end{tabular}
\end{center}
\label{tab4}
\end{table}

In this appendix,
we report more results of quantum chemistry calculations for ammonia molecules encapsulated in the $(6,5)$ CNT.
First of all, we show some characteristics of the ammonia molecule restricted to one dimension, but without CNT (Table~\ref{tab1})
and then inside the CNT (Table~\ref{tab2})
as they follow from semi-empirical methods~\cite{Christensen2016}.
Comparing numbers, we notice that the presence of the CNT results only in small changes.
This allows us to restrict ourselves
to less computer demanding case of the ammonia molecule without CNT
while performing quantum chemistry calculations
at the Hartee-Fock level (Table~\ref{tab3}) or the density-functional-theory level (Table~\ref{tab4}).

Similarly to the previous studies for water \cite{Ma2017,Druchok2023},
we observe rather scattered outcomes for the charges of nitrogen and hydrogen atoms of NH$_3$,
see the second and third rows in Tables~\ref{tab1} -- \ref{tab4}.
Most importantly and again similarly to the previous studies for water \cite{Druchok2023},
not all sets of charges $q_{\rm N}$ and $q_{\rm H}$ being plugged in MD simulations
yield two quasiphase transitions between three quasiphases.
For example,  $q_{\rm N}$ and $q_{\rm H}=-q_{\rm N}/3$ optimized at the AM1 level (Tables~\ref{tab1} and \ref{tab2})
do not reproduce a temperature-driven orientational quasiordering in MD simulations.

\appendix
\section*{Appendix~B: Mulliken population analysis for MD model for ammonia}
\renewcommand{\theequation}{B.\arabic{equation}}
\setcounter{equation}{0}

\begin{figure}
\begin{center}
\includegraphics[clip=true,width=0.9\columnwidth]{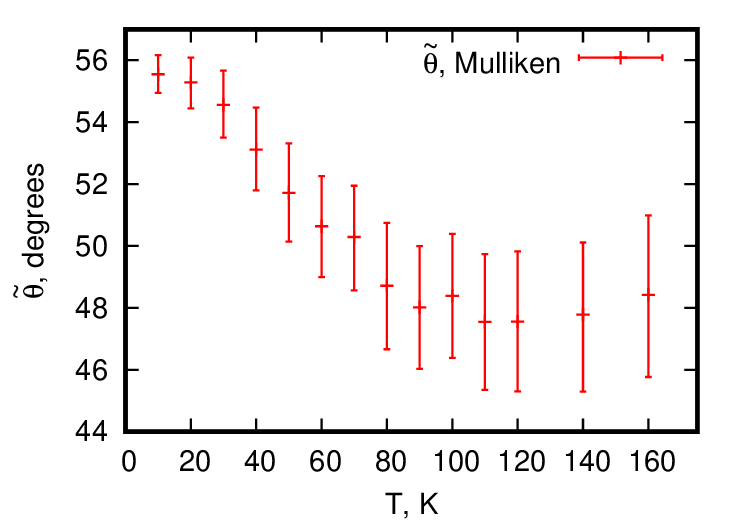}
\includegraphics[clip=true,width=0.9\columnwidth]{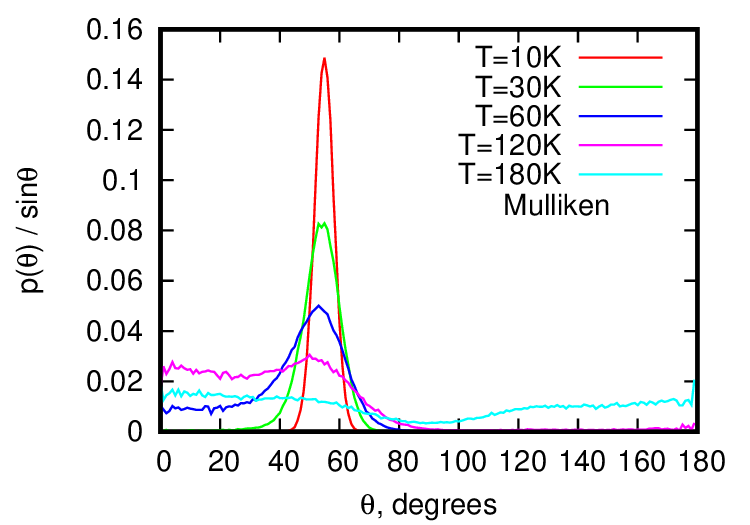}
\caption{The same as in Fig.~\ref{fig02}, however, for the ammonia molecule model with Mulliken charges.
$p(\theta)/\sin\theta$ has maximum at $\theta\approx56^\circ$ for $T=10$~K.}
\label{fig07}
\end{center}
\end{figure}

\begin{figure}
\begin{center}
\includegraphics[clip=true,width=0.9\columnwidth]{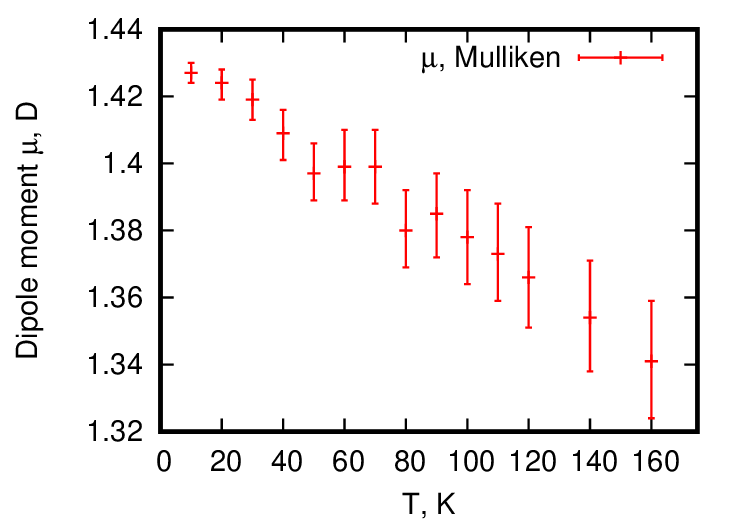}
\includegraphics[clip=true,width=0.9\columnwidth]{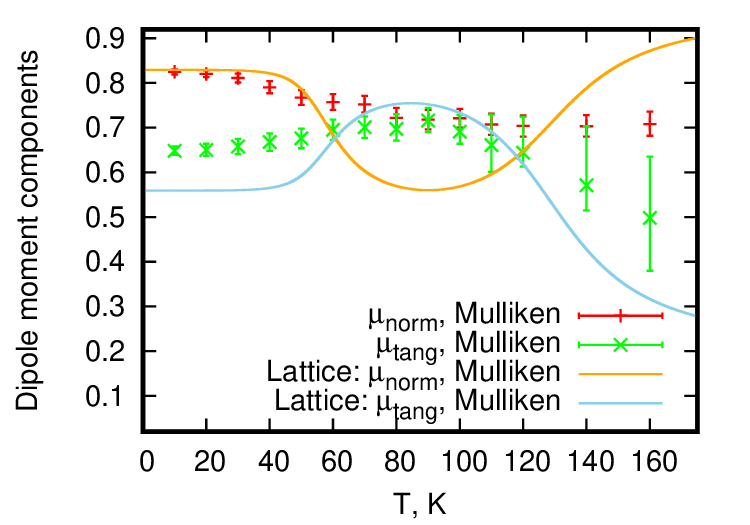}
\caption{The same as in Fig.~\ref{fig03}, however, for the ammonia molecule model with Mulliken charges.
Solid curves in the lower panel correspond to the lattice-model results, $N=10$ with $R=0.03$
for $\mu_\parallel$ (blue) and $\vert\mu_\perp\vert$ (orange).}
\label{fig08}
\end{center}
\end{figure}

\begin{figure}
\begin{center}
\includegraphics[clip=true,width=0.9\columnwidth]{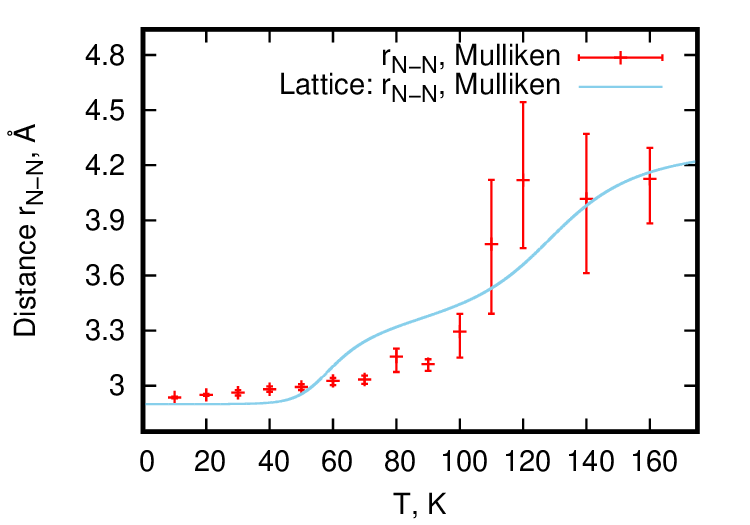}
\caption{The same as in Fig.~\ref{fig04}, however, for the ammonia molecule model with Mulliken charges.
Solid curve corresponds to the lattice-model result, $N=10$ with $R=0.03$
and $L(0)/(N-1)=2.90$~{\AA}.}
\label{fig09}
\end{center}
\end{figure}

In this appendix,
we report MD results for the ammonia molecule model with the Mulliken charges.
Namely,
Figs.~\ref{fig07}, \ref{fig08}, and \ref{fig09}
(Mulliken charges)
correspond to
Figs.~\ref{fig02}, \ref{fig03}, and \ref{fig04}
(L\"{o}wdin charges).
All shown dependences are qualitatively the same,
however,
the temperature range for the intermediate phase is shifted to higher temperatures and is $90\ldots100$~K.
In Figs.~\ref{fig08} and \ref{fig09} we also show by solid curves the lattice-model predictions,
see Sec.~\ref{sec3},
considering as an example $N=10$.
To catch the change in atomic charges we have decreased $R$ (in comparison with $R=0.07$ we used for the L\"{o}wdin charges) and set now $R=0.03$
(solid curves).
Moreover, to get the average length in angstroms (Fig.~\ref{fig09}), we set $L(0)/(N-1)=2.90$~{\AA}.
As can be seen from Figs.~\ref{fig07}, \ref{fig08}, and \ref{fig09},
overall, the lattice-model calculations again agree with the MD simulations,
although the minimum or maximum of $\mu_\parallel(T)$ (blue) and $\vert\mu_\perp\vert(T)$ (orange) are sharper than their MD counterparts
(compare to green and red symbols, respectively).

\appendix
\section*{Appendix~C: Reference data for water}
\renewcommand{\theequation}{C.\arabic{equation}}
\setcounter{equation}{0}

\begin{figure}
\begin{center}
\includegraphics[clip=true,width=0.9\columnwidth]{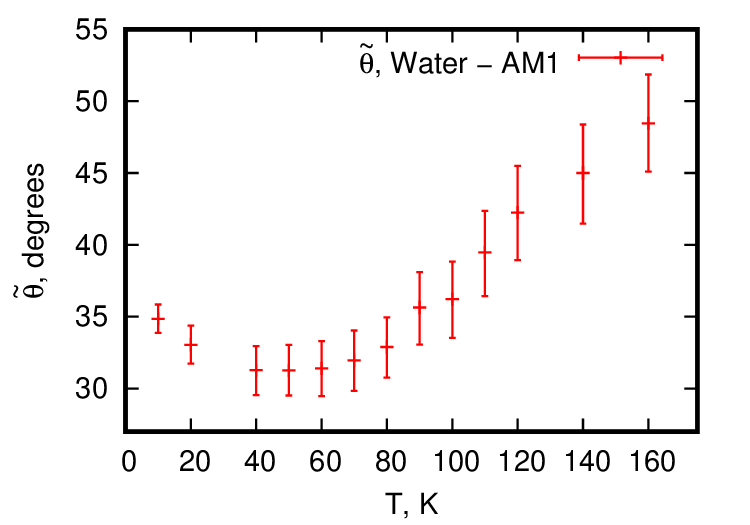}
\includegraphics[clip=true,width=0.9\columnwidth]{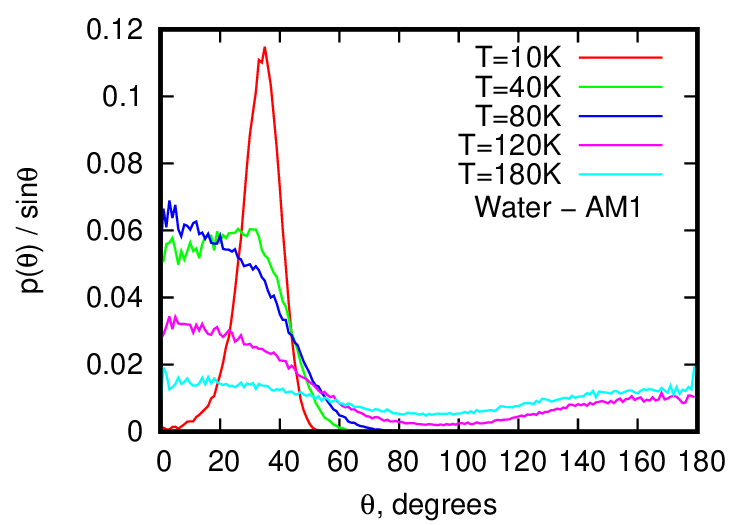}
\caption{Angles between CNT axis and dipole moments of water molecules,
cf. Figs.~\ref{fig02} and \ref{fig07} for the ammonia case.
$p(\theta)/\sin\theta$ has maximum at $\theta\approx35^\circ$ for $T=10$~K.}
\label{fig10}
\end{center}
\end{figure}

\begin{figure}
\begin{center}
\includegraphics[clip=true,width=0.9\columnwidth]{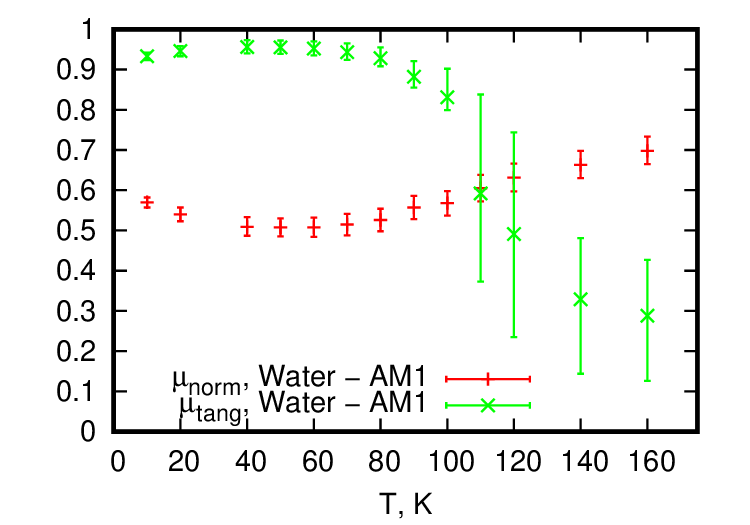}
\caption{$\mu_{\rm norm}$ (red) and $\mu_{\rm tang}$ (green) for water molecule,
cf. the bottom panels of Fig.~\ref{fig03} and Fig.~\ref{fig08}.}
\label{fig11}
\end{center}
\end{figure}

\begin{figure}
\begin{center}
\includegraphics[clip=true,width=0.9\columnwidth]{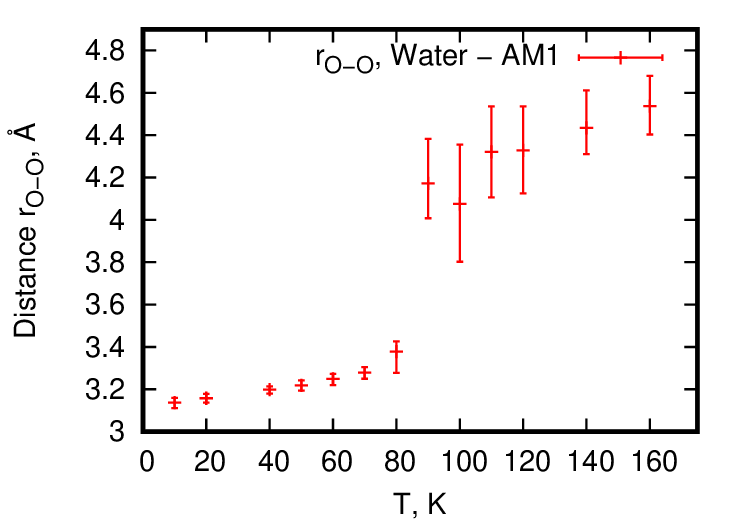}
\caption{Temperature dependence of mean distances between nearest oxygen atoms within water chain,
cf. Figs.~\ref{fig04} and \ref{fig09} for the ammonia case.}
\label{fig12}
\end{center}
\end{figure}

Water molecules encapsulated in the $(6,5)$ CNT were examined in Ref.~\cite{Druchok2023}.
However, the main focus of that paper was on the case $N=11$ water molecules encapsulated inside a CNT of $\approx40$~{\AA}.
In this appendix we show results for a longer chain of $N=35$ molecules H$_2$O in correspondence to the ammonia case.
Thus,
we
considered the $(6,5)$ CNT of the length $\approx170$~{\AA} which contains $N=35$ water molecules H$_2$O,
i.e., $4.86$~{\AA} per molecule (while $3.64$~{\AA} per molecule in Ref.~\cite{Druchok2023}),
used the SPC/E water model with the charges $q_{\rm O}=-0.4348e$ and $q_{\rm H}=0.2174e$ optimized at the AM1 level \cite{Druchok2023},
and carried out MD simulations arriving at results reported in Figs.~\ref{fig10}, \ref{fig11}, and \ref{fig12}.
Overall, the obtained results are similar to the ones for a shorter chain \cite{Druchok2023} and to those of Ref.~\cite{Ma2017}.
Note, however, that according to Ref.~\cite{Ma2017},
the most probable angle of water dipoles relative to the nanotube axis in the hydrogen-bonded quasiphase is $31^\circ$,
whereas we obtained almost $35^\circ$, Fig.~\ref{fig10}.
This may be related
to difference in the precise values of water oxygen and hydrogen charges
or
to specific water molecule model used in simulations, i.e., the TIP3P \cite{Ma2017} or SPC/E \cite{Druchok2023} water models.
On the other hand,
comparing MD simulation data in Fig.~\ref{fig03} (and Fig.~\ref{fig08}) to the ones in Fig.~\ref{fig11},
we conclude that $N=35$ ammonia and water molecules in the $(6,5)$ CNT of the length $\approx170$~{\AA} show quite similar behavior:
In both cases a quasiphase with dipoles tending to aling along the CNT axis emerges at intermediate temperatures.


\medskip

\begin{thebibliography}{99}

\bibitem{Cambre2010}
S.~Cambr\'{e}, B.~Schoeters, S.~Luyckx, E.~Goovaerts, and W.~Wenseleers,
\href{https://doi.org/10.1103/PhysRevLett.104.207401}
{Phys. Rev. Lett. {\bf 104}, 207401 (2010)}.

\bibitem{Ma2017}
X.~Ma, S.~Cambr\'{e}, W.~Wenseleers, S.~K.~Doorn, and H.~Htoon,
\href{https://doi.org/10.1103/PhysRevLett.118.027402}
{Phys. Rev. Lett. {\bf 118}, 027402 (2017)}.

\bibitem{Sahoo2021}
T.~Sahoo, T.~Serwatka, and P.-N.~Roy,
\href{https://doi.org/10.1063/5.0053051}
{J. Chem. Phys. {\bf 154}, 244305 (2021)}.

\bibitem{Serwatka2022a}
T.~Serwatka and P.-N.~Roy,
\href{https://doi.org/10.1063/5.0078770}
{J. Chem. Phys. {\bf 156}, 044116 (2022)}.

\bibitem{Serwatka2022b}
T.~Serwatka and P.-N.~Roy,
\href{https://doi.org/10.1063/5.0131149}
{J. Chem. Phys. {\bf 157}, 234301 (2022)}.

\bibitem{Serwatka2023}
T.~Serwatka, R.~G.~Melko, A.~Burkov, and P.-N.~Roy,
\href{https://doi.org/10.1103/PhysRevLett.130.026201}
{Phys. Rev. Lett. {\bf 130}, 026201 (2023)}.

\bibitem{Druchok2023}
M.~Druchok, V.~Krasnov, T.~Krokhmalskii, T.~Cardoso~e~Bufalo, S.~M.~de~Souza, O.~Rojas, and O.~Derzhko,
\href{https://doi.org/10.1063/5.0133720}
{J. Chem. Phys. {\bf 158}, 104304 (2023)}.

\bibitem{Bauer2007}
S.~H.~Bauer,
\href{https://doi.org/10.1007/s11224-007-9220-8}
{Structural Chemistry {\bf 18}, 959 (2007)}.

\bibitem{Eckl2008}
B.~Eckl, J.~Vrabec, and H.~Hasse,
\href{https://doi.org/10.1080/00268970802112137}
{Molecular Physics {\bf 106}, 1039 (2008)}.

\bibitem{Jorgensen1996}
W.~L.~Jorgensen, D.~S.~Maxwell, and J.~Tirado-Rives,
\href{https://doi.org/10.1021/ja9621760}
{J.~Am. Chem. Soc. {\bf 118}, 11225 (1996)}.

\bibitem{Schmidt1993}
M.~W.~Schmidt, K.~K.~Baldridge, J.~A.~Boatz, S.~T.~Elbert, M.~S.~Gordon, J.~J.~Jensen, S.~Koseki, N.~Matsunaga, K.~A.~Nguyen, S.~Su, T.~L.~Windus, M.~Dupuis, and J.~A.~Montgomery,
\href{https://doi.org/10.1002/jcc.540141112}
{J.~Comput. Chem. {\bf 14}, 1347 (1993)}.

\bibitem{Rizzo1999}
R.~C.~Rizzo and W.~L.~Jorgensen,
\href{https://doi.org/10.1021/ja984106u}
{J.~Am. Chem. Soc. {\bf 121}, 4827 (1999)}.

\bibitem{Mann2003}
D.~J.~Mann and M.~D.~Halls,
\href{https://doi.org/10.1103/PhysRevLett.90.195503}
{Phys. Rev. Lett. {\bf 90}, 195503 (2003)}.

\bibitem{Ertural2019}
C.~Ertural, S.~Steinberg, and R.~Dronskowski,
\href{https://doi.org/10.1039/C9RA05190B}
{Royal Society Chemistry Advances {\bf 9}, 29821 (2019)}.

\bibitem{Marx2009}
D.~Marx and J.~Hutter,
{\it Ab Initio Molecular Dynamics: Basic Theory and Advanced Methods}
(Cambridge University Press, New York, 2009).

\bibitem{Todorov2006}
I.~T.~Todorov, W.~Smith, K.~Trachenko, and M.~T.~Dove,
\href{https://doi.org/10.1039/B517931A}
{Journal of Materials Chemistry {\bf 16}, 1911 (2006)}.

\bibitem{Huang2006}
L.-L.~Huang, L.-Z.~Zhang, Q.~Shao, J.~Wang, L.-H.~Lu, X.-H.~Lu, S.-Y.~Jiang, and W.-F.~Shen,
\href{https://doi.org/10.1021/jp064676d}
{J. Phys. Chem. B {\bf 110}, 25761 (2006)}.

\bibitem{Druchok2017}
M.~Druchok and M.~Holovko,
\href{https://doi.org/10.1016/j.molliq.2016.09.093}
{J. Mol. Liq. {\bf 228}, 208 (2017)}.

\bibitem{Druchok2018}
M.~Druchok and M.~Luk\v{s}i\v{c},
\href{https://doi.org/10.1016/j.molliq.2017.11.107}
{J. Mol. Liq. {\bf 270}, 203 (2018)}.

\bibitem{Druchok2019}
M.~Druchok and M.~Luk\v{s}i\v{c},
\href{10.1016/j.molliq.2019.111287}
{J. Mol. Liq. {\bf 291}, 111287 (2019)}.

\bibitem{Yang2020}
F.~Yang, M.~Wang, D.~Zhang, J.~Yang, M.~Zheng, and Y.~Li,
\href{10.1021/acs.chemrev.9b00835}
{Chem. Rev. {\bf 120}, 2693 (2020)}.

\bibitem{Birmajer2016}
D.~Birmajer, J.~B.~Gil, and  M.~D.~Weiner,
Journal of Integer Sequences {\bf 19}, Article 16.1.3 (2016).

\bibitem{Sloane1964}
A120925: Number of ternary words on ${0,1,2}$ having no isolated $0$'s, http://oeis.org/A120925,
in: N.~J.~A.~Sloane, The On-Line Encyclopedia of Integer Sequences, http://oeis.org.

\bibitem{Hall1967}
M.~Hall,~Jr.
{\it Combinatorial Theory}
(Blaisdell Publishing Company, Waltham (Massachusetts), 1967).

\bibitem{Galitski1981}
V.~M.~Galitski, B.~M.~Karnakov, and V.~I.~Kogan,
{\it Problems in Quantum Mechanics}
(Nauka, Moscow, 1981)
(in Russian).

\bibitem{Badger1929}
R.~M.~Badger and C.~H.~Cartwright,
\href{https://doi.org/10.1103/PhysRev.33.692}
{Phys. Rev. {\bf 33}, 692 (1929)}.

\bibitem{David1996}
C.~W.~David,
\href{https://doi.org/10.1021/ed073p46}
{Journal of Chemical Education {\bf 73}, 46 (1996)}.

\bibitem{Christensen2016}
A.~S.~Christensen, T.~Kuba\v{r}, Q.~Cui, and M.~Elstner,
\href{https://doi.org/10.1021/acs.chemrev.5b00584}
{Chem. Rev. {\bf 116}, 5301 (2016)}.

\end{thebibliography}
\end{document}